\documentclass[a4paper,11pt]{article}
\usepackage{jheppub}
\usepackage{multirow}
\usepackage{booktabs}
\usepackage{amsmath,amssymb,bm,braket,psfrag,colortbl,booktabs,latexsym}

\usepackage{color}

\newcommand{\nn}{\nonumber}

\def\as{\alpha_s}
\def\nno{\nonumber}

\begin{document}


\title{ Soft gluon resummation in the signal-background interference process of $gg(\to h^*) \to ZZ$ }

\author[a,b]{Chong Sheng Li,}
\author[a]{Hai Tao Li,}
\author[c]{Ding Yu Shao}
\author[d]{and Jian Wang}

\affiliation[a]{School of Physics and State Key Laboratory of Nuclear Physics and Technology,
Peking University, Beijing 100871, China}
\affiliation[b]{Center for High Energy Physics, Peking University, Beijing 100871, China}
\affiliation[c]{Albert Einstein Center for Fundamental Physics, Institut f\"ur Theoretische Physik
Universit\"at Bern, Sidlerstrasse 5, CH-3012 Bern, Switzerland}
\affiliation[d]{PRISMA Cluster of Excellence $\&$ Mainz Institute for Theoretical Physics,
Johannes Gutenberg University, D-55099 Mainz, Germany }

\emailAdd{csli@pku.edu.cn}
\emailAdd{lihaitao@pku.edu.cn}
\emailAdd{shao@itp.unibe.ch}
\emailAdd{jian.wang@uni-mainz.de}

\date{\today}



\abstract
{We present  a precise theoretical prediction  for the signal-background interference process
of $gg(\to h^*) \to ZZ$, which is useful to constrain the Higgs boson decay width and to measure Higgs couplings to the SM particles.
The approximate NNLO $K$-factor is in the range of $2.05-2.45$ ($1.85-2.25$), depending on $M_{ZZ}$, at the 8 (13) TeV LHC.
And the soft gluon resummation can increase the approximate NNLO result by about $10\%$ at both the 8 TeV and 13 TeV LHC.
The theoretical uncertainties including the scale,
uncalculated multi-loop amplitudes of the background and PDF$+\alpha_s$ are roughly $\mathcal{O}(10\%)$ at ${\rm NNLL'}$.
We also confirm that the approximate $K$-factors in the interference and the pure signal processes are the same.

}

\arxivnumber{1504.02388}
\preprint{MITP/15-016}

\maketitle


\section{Introduction}

A scalar particle of a mass  about 125 GeV compatible with the standard model (SM) Higgs boson has been discovered recently at the LHC \cite{Aad:2012tfa,Chatrchyan:2012ufa}.
It is necessary to pin down its various quantum numbers and couplings in order to determine its identity.
The total decay width of the Higgs boson is an important variable that would appear in all the global fitting procedures.
If it is measured to be the value predicted by the SM, then the confidence to consider this particle as the SM Higgs boson is increased.
On the other hand, if it is larger than the value predicted by the SM, there would be new decay channels for this Higgs boson, e.g., invisible decay channels,
or the couplings between the Higgs boson with SM particles should be modified.
A precise measurement of the total width may open another window on new physics.

However, given that the width of the Higgs boson ($\sim 4$ MeV)
is much smaller than the energy resolution of the detector ($\sim 1$ GeV) ,
it is impossible to precisely measure the line shape and thus the total width of the Higgs boson at a hadron collider.
And one can not obtain the total decay width from global fitting of various on-shell production and decay channels \cite{Caola:2013yja}.
Taking the golden channel $gg\to h \to ZZ$ as an example, the cross section can be expressed as
\begin{equation}\label{eqs:xsec}
    \sigma \sim \frac{g_{ggh}^2 g_{hZZ}^2}{m_h\Gamma_h},
\end{equation}
where $g_{ggh}$ and $g_{hZZ}$ denote the couplings between the Higgs boson and other SM particles,
and $\Gamma_h$ is the total decay width of the Higgs boson.
Here we have used the narrow width approximation for the on-shell Higgs boson production and decay.
It is obvious that a simultaneous rescaling of couplings and width would result in the same cross section,
which means that there is no way to get independent information on the couplings or the width from only
these kinds of measurements.

\begin{figure}\center
  \includegraphics[width=0.95\linewidth]{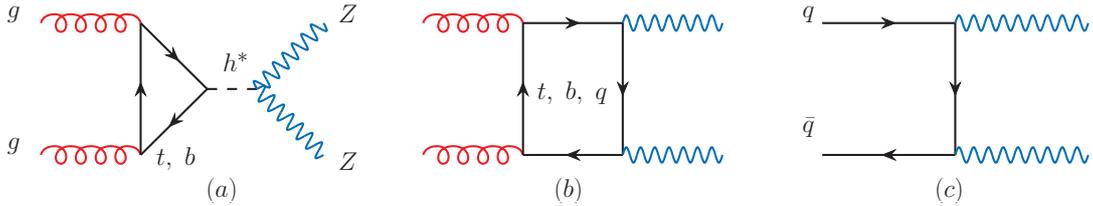}\\
  \caption{The leading order Feynman diagrams for the signal and background.
  Diagrams $(a)$ and $(b)$ denote the amplitude of signal and background, respectively. Their interference contributes to the signal.
  Diagrams $(c)$ represents the dominant SM background.   }
  \label{eps:lo}
\end{figure}

Other proposals have been presented in the literatures so far to bound the total width, which  depend on additional model or mass resolution assumptions~\cite{Martin:2012xc,Martin:2013ula,Dixon:2013haa,Kauer:2012hd,Caola:2013yja,Campbell:2013una}.
For example, it is assumed that the Higgs coupling to a $W$ or $Z$ boson pair
is not much larger than in the SM \cite{Dobrescu:2012td},
or the Higgs couples only to the SM particles \cite{Barger:2012hv}.
A method to bound the Higgs boson width through apparent mass shift  due to interference between the Higgs resonance in gluon fusion and the continuum background amplitude for $gg\to h \to \gamma \gamma $
was investigated \cite{Martin:2012xc,Martin:2013ula},
but the experimental mass resolution is modeled by a Gaussian distribution for simplicity \cite{Dixon:2013haa}.
A recent proposal makes use of the information of the cross section at the non-resonance region
where the final states have an invariant mass larger than 125 GeV \cite{Kauer:2012hd,Caola:2013yja,Campbell:2013una}.
Actually, looking into the non-resonance region,
where $\hat{s}\approx M_{ZZ}^2 \gg m_h^2$ ($M_{ZZ}$ is the invariant mass of the $Z$ boson pair),
brings two changes to the formula in eq.(\ref{eqs:xsec}).
First, the narrow width approximation is not applicable any more and thus $\Gamma_h$ in the propagator can be neglected.
And because of the $Z$ boson pair and top quark pair threshold effects,
the cross section away from the Higgs threshold is not negligible small.
Second, the interference process
between the signal (figure \ref{eps:lo}(a)) and background (figure \ref{eps:lo}(b)) becomes important.
And the cross section at the non-resonance region is given by
\begin{equation}\label{eqs:xsec2}
    \frac{d\sigma}{dM_{ZZ}^2} \sim \frac{g_{ggh}^2 g_{hZZ}^2}{(M_{ZZ}^2-m_h^2)^2}
    +\frac{g_{ggh} g_{hZZ} }{(M_{ZZ}^2-m_h^2)M_{ZZ}^2},
\end{equation}
where the first and second term arise from the pure signal and signal-background interference processes, respectively.
From eq.(\ref{eqs:xsec2}), we see that the cross section in the non-resonance region is only sensitive to the Higgs boson couplings.
Combining the information from both the on-shell and off-shell regions provides a way to measure or bound the Higgs boson total width.

The above statement is actually also based on  assumptions. For example,
the couplings $g_{ggh}$ and $g_{hZZ}$ do not vary when changing from
on-shell to off-shell regions, or have a known dependence on $M_{ZZ}$.
And the discussion of the  limitations of the off-shell coupling measurements has been available recently~\cite{Englert:2014aca,Buschmann:2014sia,Logan:2014ppa}.
Nevertheless, with this method, the ATLAS and CMS collaborations have obtained the upper limit of
$\Gamma_h< 4.8\sim 7.7 \Gamma_h^{\rm SM}$  and $\Gamma_h< 5.4\Gamma_h^{\rm SM}$, respectively,
at a $95\%$ confidence level \cite{Khachatryan:2014iha,atlas1}.

Moreover, the off-shell Higgs production and decay has significant impact on search for new physics in addition to
the interpretation of the Higgs total width \cite{Gainer:2014hha,Azatov:2014jga,Englert:2014ffa}.
And the on-shell Higgs production can not distinguish the contributions
from the $htt$ and $hgg$ (induced by new colored particle loop) couplings  since
they would give rise to the same effective operator for a single Higgs boson on-shell production production and decay.
The off-shell Higgs production breaks this degeneracy because the $hgg$ coupling
is sensitive to the off-shellness of the Higgs boson
\footnote{Another method to break the degeneracy is boosted Higgs production.}.
Therefore, from the off-shell Higgs production, one can obtain a constraint on the $htt$ coupling with the similar accuracy to that from
the $pp\to t\bar{t}h$ production \cite{Cacciapaglia:2014rla}.

Though the theoretical prediction for the pure signal process is known up to next-to-leading order (NLO)
and next-to-next-to-leading logarithmic (NNLL) level
with the finite top quark mass \cite{Dittmaier:2011ti,Heinemeyer:2013tqa,Passarino:2013bha},
the interference process is predicted only at leading order (LO)\cite{Kauer:2012hd,Kauer:2013qba,Campbell:2013una}.
Some recent papers provide the QCD radiative corrections to the production of $gg\to ZZ$ via a top quark loop in the heavy top quark limit \cite{Melnikov:2015laa} and via a massless quark loop \cite{Caola:2015ila,vonManteuffel:2015msa}.
A high-order soft-collinear approximation has been obtained
for the signal-background interference of a heavy Higgs boson (600 GeV)
production and decay into $WW$ in ref.\cite{Bonvini:2013jha},
which shows that the $K$-factors of the approximate NLO and next-to-next-to-leading order (NNLO) corrections are very sizable,
and the theoretical uncertainties are still large, about $9\%$ at approximate NNLO.
This result implies that the soft gluon effects are important, which should be resummed up to all order, and provide more reliable theoretical predictions.

There are several differences between a heavy Higgs boson (600 GeV)
production with decay into $WW$ and the SM Higgs boson (125 GeV) off-shell production with decay into $ZZ$.
First, the SM Higgs boson is relative light, and  the contribution
from the signal-background interference is different, compared to the heavy Higgs boson. More explicitly, the interference cross section
for a heavy Higgs boson is positive while the one for the SM Higgs boson is negative.
Second, the interference for a heavy Higgs boson is most significant around its mass threshold, while the interference for the SM Higgs boson dominates around $M_{ZZ}=200$ GeV.
Third, there are additional contributions at high orders in the SM Higgs boson (125 GeV) off-shell production and decay into $ZZ$
because of the contributions from the $ggZ$ triangle loop diagrams; see the diagram \ref{eps:nlo}$(c)$.
Last, searching for $WW$ final states requires an additional jet veto applied to suppress the large $t\bar{t}$ background.
This would induce another kind of large logarithms that need also to be resummed to all order \cite{Moult:2014pja}.

Note that in the large invariant mass region, the $Z$ bosons are significantly boosted.
The dominant contribution comes from the longitude component of the $Z$ boson, which is similar to a Higgs boson.
Therefore, it is expected the impact from soft gluon resummation in this process is similar
to that in the Higgs boson pair production, which we have studied earlier  \cite{Shao:2013bz}.

This paper is organized as follows.
In section \ref{sec:resum}, we describe the resummation formalism in this process briefly.
We then investigate the NLO and NNLO expansions from the resummation formalism in section \ref{sec:nnlo}.
In section \ref{sec:numerical}, we discuss the numerical results including the
invariant mass distributions and the theoretical uncertainties.
The conclusions are given in section \ref{sec:conclusion}.

\section{Resummation formalism}
\label{sec:resum}
In this paper, we concentrate on the interference process, i.e.
the interference between diagrams \ref{eps:lo}$(a)$ and \ref{eps:lo}$(b)$,
which contributes to the signal.
The amplitude squared of diagram \ref{eps:lo}$(a)$  is taken to be the pure signal, which has been calculated very precisely.
The amplitude squared of the diagram \ref{eps:lo}$(b)$ is considered as background.
However, this kind of background is much less than the one coming from diagram \ref{eps:lo}$(c)$.
The amplitudes of the diagrams in figure \ref{eps:lo} have been computed at LO, and can be found in
$\tt{gg2VV}$ \cite{Kauer:2012hd,Kauer:2013qba} and $\tt{MCFM}$ \cite{Campbell:2013una}.

\begin{figure}\center
  \includegraphics[width=0.95\linewidth]{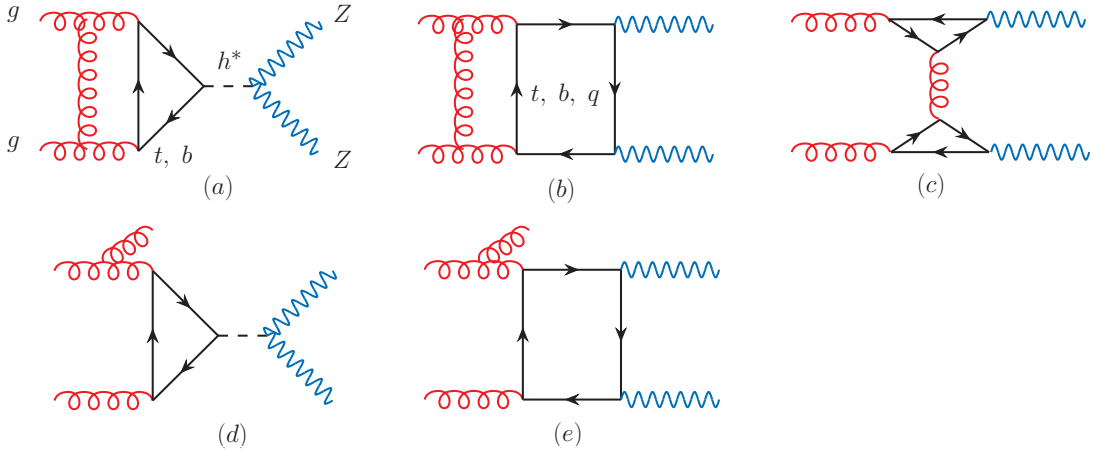}\\
  \caption{The next-to-leading order Feynman diagrams for the interference process.
  Diagrams $(a)$ and $(b)$ denote the virtual correction while diagrams $(e)$ and $(f)$ denote the real correction.
  Diagram $(c)$ represents a new contribution, compared with $gg\to WW$, to the background that could interference with the signal.   }
  \label{eps:nlo}
\end{figure}

More precise predictions require the calculation of the diagrams in figure \ref{eps:nlo}.
This has not been achieved so far due to the complex two-loop diagrams with massive particles in both loops and external states.
Given that the invariant mass of the final state is large, the soft gluon contribution is expected to dominate the higher order corrections.
And this special contribution can be resummed to all orders in $\alpha_s$.
In this section, we describe the neccessary resummation formalism briefly.
A more detailed discussion of the factorization and resummation formalism can be found in the
single Higgs \cite{Ahrens:2008nc} or the double Higgs \cite{Shao:2013bz} productions  at the LHC.

In the off-shell Higgs production and decay to an on-shell $Z$ boson pair,
the invariant mass of the $Z$ boson pair $M_{ZZ}$ is so large that additional real emissions are strongly constrained.
The imbalance between the virtual and real corrections induces large logarithms of the form $\ln^n (1-z)$, where
the partonic threshold variable $z$ is defined as
\begin{equation}\label{eqs:partonic_threshold}
    z\equiv \frac{M_{ZZ}^2}{\hat{s}}
\end{equation}
with the variable $\sqrt{\hat{s}}$ being the partonic center-of-mass energy.
The hadronic threshold region is defined as
\begin{equation}\label{eqs:threshold}
    \tau\equiv \frac{M_{ZZ}^2}{s} \to 1
\end{equation}
with $\sqrt{s}$ being the collider energy.
In the threshold region, the differential cross section can be written as
\begin{align}
\frac{ d\sigma\left(M_{ZZ}\right)}{dM_{ZZ}^2}
 =\frac{1}{s} \int_\tau^1 \frac{d z}{z} f\!\!f_{gg}\left(\frac{\tau}{z},\mu_f \right)
  d\hat{\sigma}_B \left(M_{ZZ}\right)  C(z,M_{ZZ},\mu_f),
\end{align}
where $f\!\!f_{gg}$ is the luminosity of the initial-state gluons in the protons, defined as
\begin{align}
f\!\!f_{gg}(y,\mu) = \int_{y}^1 \frac{dx}{x} f_{g/N_1}(x,\mu) f_{g/N_2}(y/x,\mu).
\end{align}
And the partonic differential cross section is given by
\begin{align}
d\hat{\sigma}_B(M_{ZZ})  = \frac{1}{2\hat{s}}
\overline{\left| \mathcal{M} \right|_B^2}  \,d \mathbf{\Phi}_2.
\end{align}
Here, $\overline{\left| \mathcal{M} \right|_B^2}$ is the LO color and spin sum (averaged) amplitude squared,
and $d \mathbf{\Phi}_2$ denotes the two-body phase space of the $Z$ boson pair.
In addition, $C(z,M_{ZZ},\mu_f)$ represents the hard scattering kernel.
After applying the similar derivation procedure for Drell-Yan process~\cite{Becher:2007ty}
in the framework of
soft-collinear effective theory (SCET) \cite{Bauer:2000ew,Bauer:2000yr,Bauer:2001ct,Bauer:2001yt,Becher:2006nr},
it could be factorized to two parts, i.e., the hard function and the soft function,
\begin{align}
C(z,M_{ZZ},\mu_f) = \mathcal{H}(M_{ZZ},\mu_f) \mathcal{S}(\sqrt{\hat{s}}(1-z),\mu_f).
\end{align}

The hard functions for the signal and interference processes are given respectively by
\begin{align}
\mathcal{H}_{\rm sig}(M_{ZZ},\mu) &=\left| C_h(i\,M_{ZZ},\mu)  \right|^2, \nno \\
\mathcal{H}_{\rm int}(M_{ZZ},\mu) &= \mathcal{R}e\left[ C_h(i\,M_{ZZ},\mu) C_{VV}^{\ast}(i\,M_{ZZ},\mu) \right],
\end{align}
where the hard Wilson coefficient $C_a(iM_{ZZ},\mu)$ is obtained by matching the gluon operator from QCD to SCET \cite{Ahrens:2008nc}. Here the index $a=h$ and $VV$ accounts for the process of $gg\to h^*\to ZZ$ and $gg\to ZZ$, respectively.
The renormalization group (RG) equation for the hard Wilson coefficient is
\begin{align}
\frac{d C_{a}(i M_{ZZ},\mu)}{d\ln \mu} = \left[ \Gamma_{\rm cusp}^A(\as) \ln\frac{-M_{ZZ}^2}{\mu^2} + \gamma^{gg}(\as) \right] C_{a}(iM_{ZZ},\mu),
\end{align}
and the corresponding solution (evolved to the factorization scale) is
\begin{align}\label{eqs:hardWilson}
C_{a}(iM_{ZZ},\mu_f) =
\exp\left[2 S(\mu_h,\mu_f) - a_\Gamma(\mu_h,\mu_f)\ln\frac{-M_{ZZ}^2}{\mu_f^2} - a_{\gamma^{gg}}(\mu_h,\mu_f)\right] C_{a}(iM_{ZZ},\mu_h),
\end{align}
where the intrinsic hard scale $\mu_h$ is chosen as $\mu_h^2=-M_{ZZ}^2$ and the $\pi^2$-enhanced terms are resummed to all order in $\alpha_s$ by RG evolution from $\mu_h^2=-M_{ZZ}^2$ to positive values of $\mu_f^2$~\cite{Ahrens:2008nc}.
$S(\nu,\mu)$ and $a_{\gamma}(\nu,\mu)$ are functions respectively
of the anomalous dimensions $\Gamma_{\rm cusp}^A$ and $\gamma$,
as defined in \cite{Becher:2006nr}, and we do not write them explicitly here.
At the fixed orders, the hard Wilson coefficients have perturbative expressions in series of the strong coupling,
\begin{align}
C_{a}(iM_{ZZ},\mu) = 1 + \sum_{n=1}^{\infty} c_{a}^{(n)}(L)\left(\frac{\as}{4\pi}\right)^n,
\end{align}
where $L=\ln\left(-M_{ZZ}^2/\mu^2\right)$.
In the matching procedure for $gg\to h^*\to ZZ$,  the Wilson coefficient at NLO can be expressed as
\begin{align}
c_{h}^{(1)}(L) =& -C_A L^2 +  \left[\frac{\pi^2}{2}+c_1\right].
\end{align}
Here we have written the scale dependent part explicitly.
$c_1$ is scale independent and its precise value requires the calculation of the two-loop diagrams,
such as the diagram \ref{eps:nlo}$(a)$,
and can be extracted from refs.\cite{Spira:1995rr,Harlander:2005rq,Aglietti:2006tp,Ball:2013bra}.
In the large top quark mass limit, $c_1=11$.
In our numerical discussion, we have included the exact finite quark mass effect in the signal amplitude.
Up to NNLO, i.e. three-loop level, we can only make use of the result obtained in the large top quark mass limit at the moment,
and $c_{h}^{(2)}(L)$ can be approximately expressed as
\begin{align}
c_{h}^{(2)}(L) =& C_A^2 H_A(L) + C_A T_F n_f H_f(L) +  \left[\frac{\pi^2}{2}+c_1\right] H_1(L)+ c_2 ,
\end{align}
where
\begin{align}
c_2 =&\left( \frac{7451}{54} + \frac{217\pi^2}{12} + \frac{\pi^4}{8} - \frac{499}{3}\zeta_3 \right),
\end{align}
and $H_A$, $H_f$ and $H_1$ have the form
\begin{align}
H_A(L)  &= \frac{L^4}{2} + \frac{11}{9}L^3 + L^2\left( -\frac{67}{9} + \frac{\pi^2}{3} \right) + L\left( \frac{386}{27} - \frac{11\pi^2}{18} - 2\zeta_3 \right), \nno \\
H_f(L)  &= -\frac{4}{9}L^3 + \frac{20}{9}L^2 + L\left( -\frac{76}{27} + \frac{2}{9}\pi^2 \right), \nno \\
H_1(L) & = - C_A L^2 + L\left( - \frac{11}{3} C_A + \frac{4}{3} n_f T_F \right),
\end{align}
respectively.
At the moment, there are no complete results for the two-loop or three-loop virtual corrections to the background $gg\to ZZ$ amplitude.
However we solve the RG group for the hard Wilson coefficient, and get the exact scale dependent terms as
\begin{align}
c_{VV}^{(1)}(L) &= -C_A L^2 +  \delta_1 , \nno \\
c_{VV}^{(2)}(L) &= C_A^2 H_A(L) + C_A T_F n_f H_f(L) + \delta_1 H_1(L) + \delta_2,
\end{align}
where $\delta_1$ and $\delta_2$ represent the unknown scale independent terms at the two-loop and three-loop levels respectively.

The soft function describes soft interactions between all external colored particles.
It has the same form as in the single Higgs production \cite{Ahrens:2008nc}. Up to $\mathcal{O}(\as)$, it is given by
\begin{align}
\mathcal{S}\left( \sqrt{\hat{s}}(1-z),\mu \right) & =
 \delta(1-z) + \frac{\as}{\pi}\left[ \left(\frac{3}{2}L^2 + \frac{\pi^2}{4}\right) \delta(1-z) + 6D(z) \right]
\end{align}
with
\begin{align}
D(z) = \left[ \frac{1}{1-z} \ln\frac{M_{ZZ}^2(1-z)^2}{z\mu^2} \right]_+  .
\end{align}
This kind of plus distribution comes from the integration with the pole subtracted. The soft function obeys an integro-differential evolution equation, which has been shown in ref.\cite{Shao:2013bz}. Using the Laplace transformation~\cite{Ahrens:2008nc}, we can obtain the corresponding solution as
\begin{align}
\mathcal{S}\left( \sqrt{\hat{s}}(1-z),\mu_f \right) = U_s(\mu_s,\mu_f)\, \tilde{s}(\partial_\eta,\mu_s) \frac{\hat{s}^\eta(1-z)^{2\eta-1}}{\mu_s^{2\eta}} \frac{e^{-2\gamma\eta}}{\Gamma(2\eta)},
\end{align}
where the auxiliary parameter $\eta$ is defined as $\eta=2a_{\Gamma}(\mu_s,\mu_f)$, and
\begin{align}
U_s(\mu_s,\mu_f) = \exp\left[ -4S(\mu_s,\mu_f) + 2 a_{\gamma^W}(\mu_s,\mu_f)\right]
\end{align}
with $\mu_s$ being the intrinsic soft scale and usually set numerically.
We choose it according to the method in \cite{Shao:2013bz}.
The function $\tilde{s}(\partial_\eta,\mu_s)$ is the Laplace transformed soft function, which is defined as
\begin{equation}
\tilde{s}(L,\mu_s)=\int_0^{\infty}d\omega e^{-\omega/(e^{\gamma_E}\mu_s e^{L/2})}
\mathcal{S}\left( \omega ,\mu_s \right).
\end{equation}
Using the property of RG invariance of the total cross section, we could get the anomalous dimension $\gamma^W$ as
\begin{align}
\gamma^W = \frac{\beta(\as)}{\as} + \gamma^{gg} + 2 \gamma^B.
\end{align}

Combining the above components together, we obtain the resummed hard scattering coefficient for the interference process
\begin{align}
C(z,M_{ZZ},\mu_f) = & \mathcal{R}e\left[C_h(i\,M_{ZZ},\mu_f) C_{VV}^*(i\,M_{ZZ},\mu_f)\right] U(M_{ZZ},\mu_h,\mu_s,\mu_f) \nno \\
& \times \frac{z^{-\eta}}{(1-z)^{1-2\eta}} \tilde{s}\left(\ln\frac{M_{ZZ}^2(1-z)^2}{z\,\mu_s^2} + \partial_\eta, \mu_s\right) \frac{e^{-2\gamma \eta}}{\Gamma(2\eta)} ,
\end{align}
where
\begin{align}
U(M_{ZZ},\mu_h,\mu_s,\mu_f) = & \frac{\as^2(\mu_s)}{\as^2(\mu_f)}\left| \left(\frac{-M_{ZZ}^2}{\mu_h^2}\right)^{-2a_\Gamma(\mu_h,\mu_s)} \right| \nno \\
&\times \left| \exp\left[ 4S(\mu_h,\mu_s) - 2a_{\gamma^{gg}}(\mu_h,\mu_s) + 4a_{\gamma^{B}}(\mu_s,\mu_f) \right] \right|  .
\end{align}
This is the main formula in our calculation.
In practice, we would use the three-loop cusp anomalous dimension
and two-loop normal anomalous dimension,
and thus denote the precision of the resummed result as ${\rm NNLL}'$,
where the prime  means that the results are not exact ${\rm NNLL}$
due to the existence of unknown scale independent terms  in the hard function.
Notice that the $\pi^2$-enhanced terms have been resummed to NNLL order.
Since the fixed-order result is only exactly known to LO, we do not match the resummed result with fixed-order ones.

\section{NLO  and NNLO expansions}
\label{sec:nnlo}

In the resummed cross section, there are three scales, $\mu_f$, $\mu_s$ and $\mu_h$.
If we set them equal to each other, then we obtain the threshold singular contributions,
which should appear in the fixed-order calculations.
Up to NNLO, the expanded result is given by
\begin{align}\label{eqs:expansion}
C(z,M,\mu_f) = & \delta(1-z) + \frac{\as}{\pi}\left\{ \delta(1-z)\left[ \frac{15\pi^2}{8} + \frac{c_1}{4} + \frac{\delta_1}{4}\right]  + 2 C_A P'_1(z) \right\} \nno \\
& + \left(\frac{\as}{\pi}\right)^2\Big[ \, C_A^2 S_A(z) + C_A T_F n_f S_f(z)  +  S_1(z) \,\Big],
\end{align}
where the auxiliary function $P'_n(z)$ is defined as
\begin{align}
P'_n(z) = \left[ \frac{1}{1-z} \ln^n \left( \frac{M_{ZZ}^2(1-z)^2}{\mu_f^2 z} \right) \right]_+.
\end{align}
Besides, the two-loop coefficients $S_i(z)$ are defined as
\begin{align}
S_A(z) =\,&  \delta (1-z) \left[\left(\frac{7 \text{$\zeta_3$}}{2}-\frac{77 \pi ^2}{144}-\frac{1}{12}\right) L_M -\frac{55 \text{$\zeta_3$}}{72} +\frac{31 \pi^4}{288}+\frac{871 \pi ^2}{864}+\frac{607}{324}\right] \nno \\
   & +   P'_3(z) - \frac{11}{12}P'_2(z) + \left( \frac{67}{18} - L_M^2 - \pi^2 \right)P'_1(z) + \left( -\frac{101}{27} + \frac{11\pi^2}{18} + \frac{39\zeta_3}{2} \right) P'_0(z), \nno \\
   S_f(z) =\, & \delta (1-z) \left[\left(\frac{7 \pi ^2 }{36}+\frac{1}{6}\right)L_M + \frac{5 \text{$\zeta_3 $}}{18} -\frac{65 \pi^2}{216}-\frac{41}{81}\right] +\frac{1}{3} P'_2(z) -\frac{10}{9}
   P'_1(z)  \nno \\
   &+\left(\frac{28}{27}-\frac{2 \pi ^2}{9}\right) P'_0(z), \nno \\
   S_1(z) =\, & \delta(1-z) \Bigg[  -\frac{23}{96}  \left(2 c_1+2 \text{$\delta_1$}+\pi ^2\right) L_M + \frac{1}{16} \left( \delta_1 + 7\pi ^2\right) c_1   + \frac{15 \pi ^2 \text{$\delta_1$}}{32}+\frac{\text{$\delta_2$}}{16}-\frac{499 \text{$\zeta_3$}}{48} \nno \\
   & +\frac{29 \pi
   ^4}{128}+\frac{217 \pi ^2}{192}+\frac{7451}{864}
    \Bigg] + \left[ \frac{3}{2}\left( c_1 + \delta_1 \right)  + \frac{3\pi^2}{4}\right] P'_1(z),
 \end{align}
where the notation $L_M$ is defined as $L_M = \ln(M_{ZZ}^2/\mu_f^2)$.
The NLO and NNLO results obtained this way include contributions from the complete one- and two-loop virtual corrections
\footnote{This means that the actual diagrams at NLO and NNLO are of two and three loops if the LO is already of one loop.}
and soft gluon real corrections. All the scale dependent parts are process independent and have been incorporated in
various anomalous dimensions. The unknown scale independent parts are represented by $\delta_{1,2}$.

\section{Numerical results}
\label{sec:numerical}

In numerical calculation, we take the SM input
\begin{align}\label{eqs:input}
     \quad m_h & =125.7 ~{\rm GeV},\quad m_t  = 173.2 ~{\rm GeV}, \quad m_b=4.89 ~{\rm GeV},  \nn \\
      M_Z & =91.1876 ~{\rm GeV}, \quad \Gamma_Z  =2.4952 ~{\rm GeV},  \quad   M_W = 80.398~{\rm GeV}, \nn \\
      G_F&=1.16639\times 10^{-5}~ {\rm GeV}^{-2}, \quad
    \alpha(M_Z) =1/132.338.
\end{align}
The factorization and renormalization scales are set to be $M_{ZZ}$.
When discussing the scale uncertainties, we vary them from $M_{ZZ}/2$ to $2M_{ZZ}$.
We use the MSTW2008LO, NLO, NNLO PDF sets \cite{Martin:2009iq} and associated strong coupling constant to calculate the
LO, NLO and NNLO results, respectively.
We are interested in the results at the 8 TeV and 13 TeV LHC.
In our calculation, we also consider the decay of the $Z$ bosons, e.g., $ZZ\to e^+ e^- \mu^+ \mu^-$,
so the invariant mass of the four final leptons $M_{4l}=M_{ZZ}$.

The unknown NLO and NNLO results for the non-logarithmic parts in the hard function of the interference process, i.e., $\delta_1$ and $\delta_2$, are estimated
by using the analogous results in the pure signal process and varying by a factor of $\xi_{1,2}$.
More specifically, the non-logarithmic parts in the hard function of the interference process are approximated as
\begin{eqnarray}
  \delta_1 &=& \xi_1 \left( \frac{\pi^2}{2}+ c_1 \right),  \\
  \delta_2 &=& \xi_2 \, c_2.
\end{eqnarray}
The default value of $\xi_{1,2}$ is chosen to be 1. We vary $\xi_{1,2}$
from 0 to 2 to estimate the uncertainties coming from those uncalculated
non-logarithmic parts in higher order virtual corrections.

\begin{figure}\center
  \includegraphics[width=0.48\linewidth]{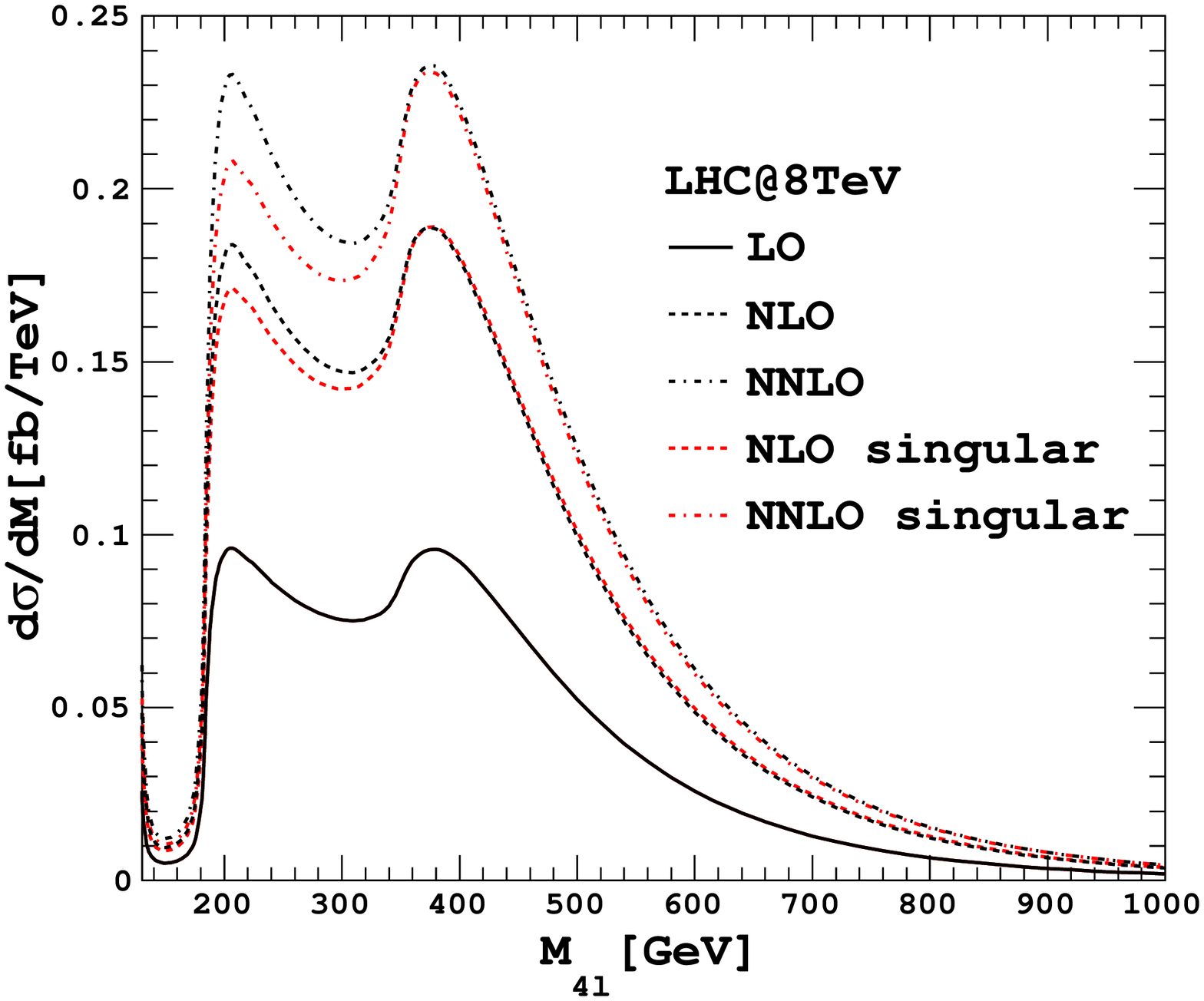}
  \includegraphics[width=0.48\linewidth]{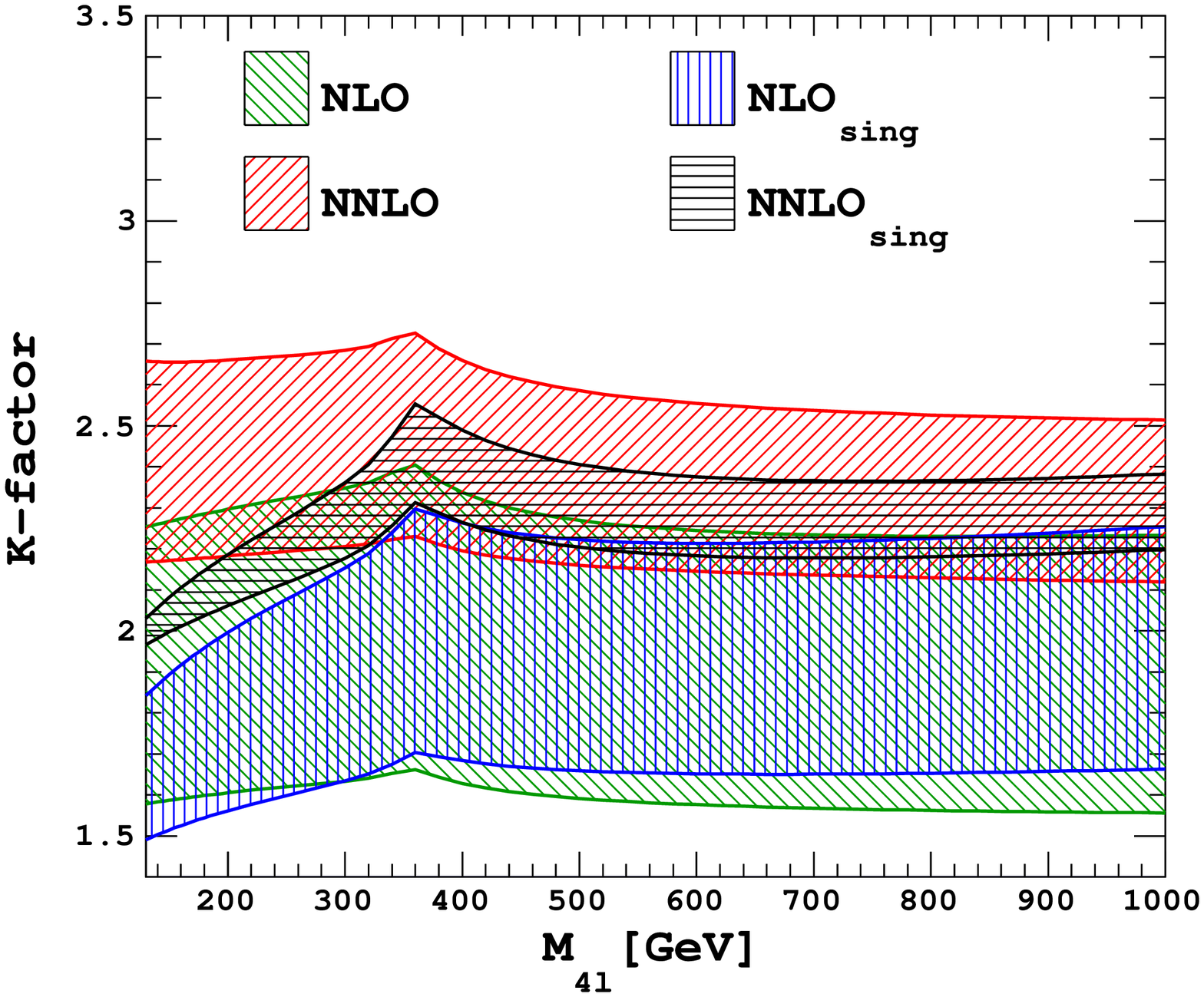}
  \\
  \caption{The cross section and $K$-factor for the pure signal processes at the 8 TeV LHC.  }
  \label{eps:signal}
\end{figure}

\begin{figure}\center
  \includegraphics[width=0.48\linewidth]{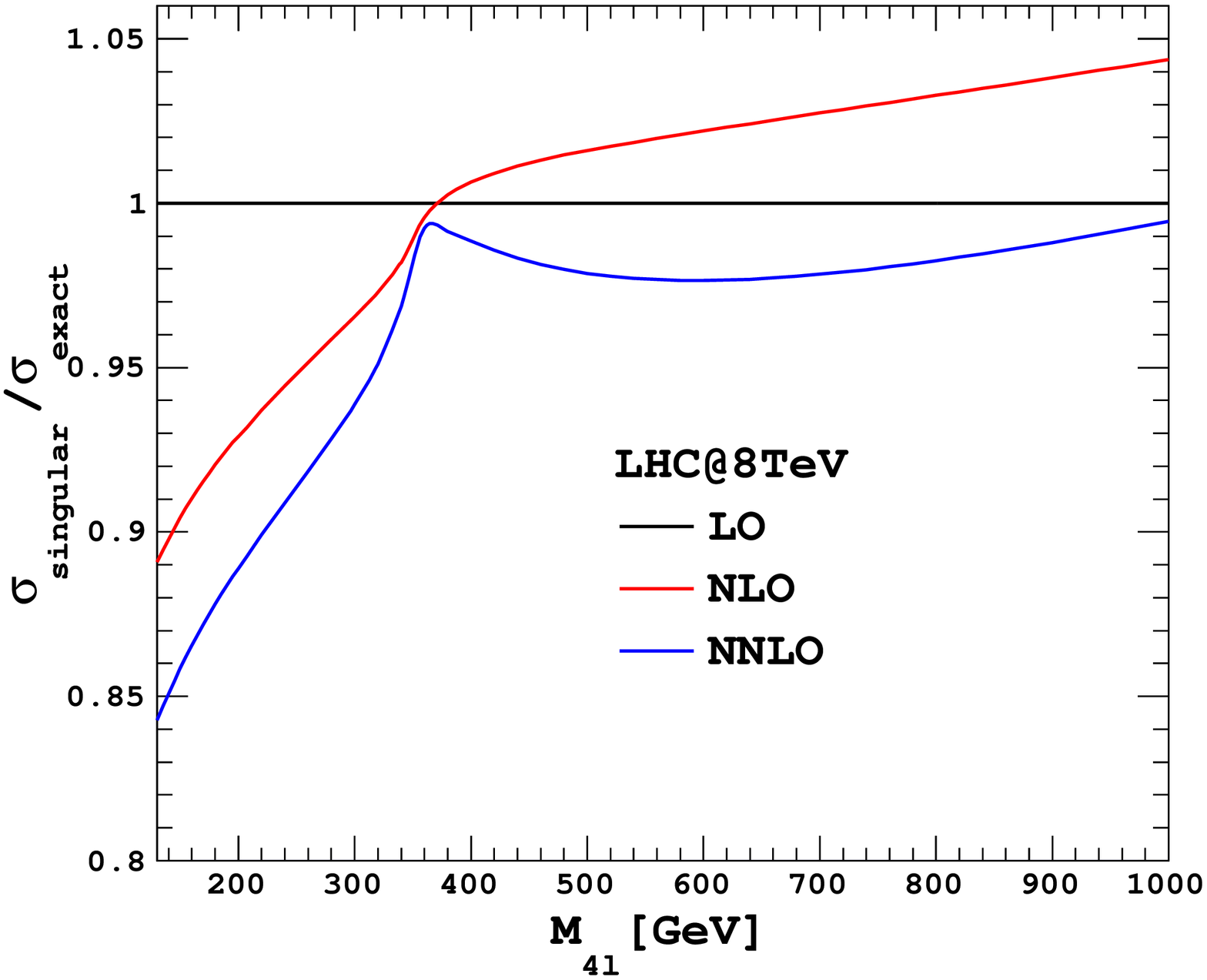}
  \includegraphics[width=0.48\linewidth]{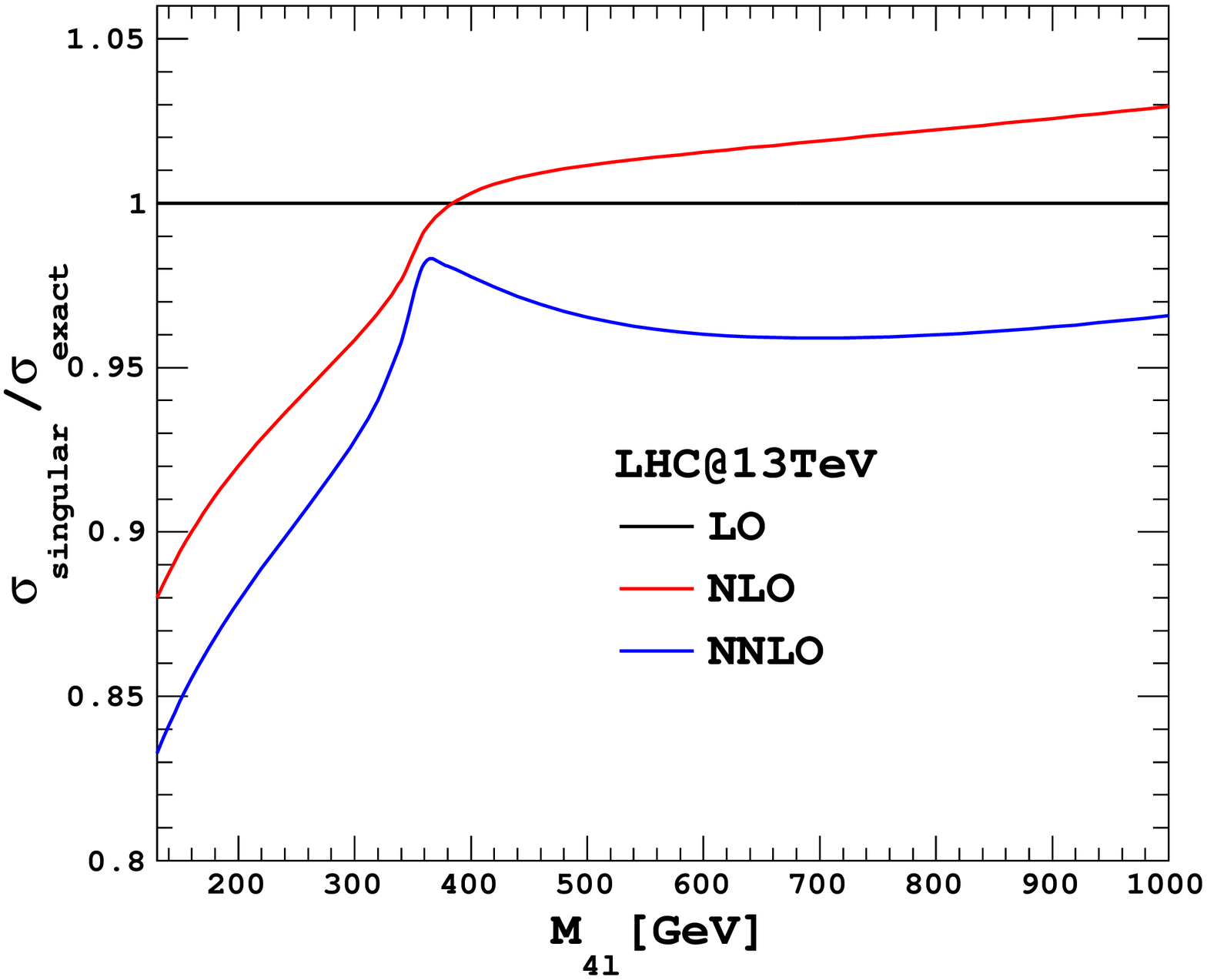}
  \\
  \caption{The fraction of the singular terms to the exact cross sections.  }
  \label{eps:domi}
\end{figure}

Before we present the resummed result for the signal, we first estimate that to what extent the resummed result is valid.
Since the pure signal process has the same initial and final state as the inference process, we take it to illustrate this point.
The LO, NLO, and NNLO singular contributions,
obtained according to eq.(\ref{eqs:expansion}) in the above section, along with the exact results,
calculated by  {\tt FehiPro}
\footnote{Both the claimed exact NLO and NNLO results contains exact top quark mass dependence only up to NLO.} \cite{Anastasiou:2005qj,Anastasiou:2009kn}, are shown in figure \ref{eps:signal}.
Their ratios are shown in  figure \ref{eps:domi}.
We first notice that the exact NLO and NNLO $K$-factors are very significant,
and almost constants over a large region of $M_{ZZ}>2M_Z$.
Then it is evident that the contribution of the singular terms dominates the higher order corrections.
The singular contribution almost coincides with the exact results for $M_{ZZ}> 2m_t$  at both NLO and NNLO;
the difference is below $2\%$ at NNLO.
In the smaller $M_{ZZ}$ region, the singular contributions are a little less than the exact results.
But the  difference is less than $4\%$ and $10\%$ at NLO and NNLO, respectively, for $M_{ZZ}> 220$ GeV,
which is the off-shell region defined in experiments \cite{Khachatryan:2014iha,atlas1}.
Moreover, the scale uncertainties of the singular terms lie in or highly overlap with those of the exact results for $M_{ZZ}> 220$ GeV,
as shown in the right plot of figure \ref{eps:signal}.
From this comparison, it is reasonable to use only the singular terms to predict the unknown higher order effects.

\begin{figure}\center
  \includegraphics[width=0.45\linewidth]{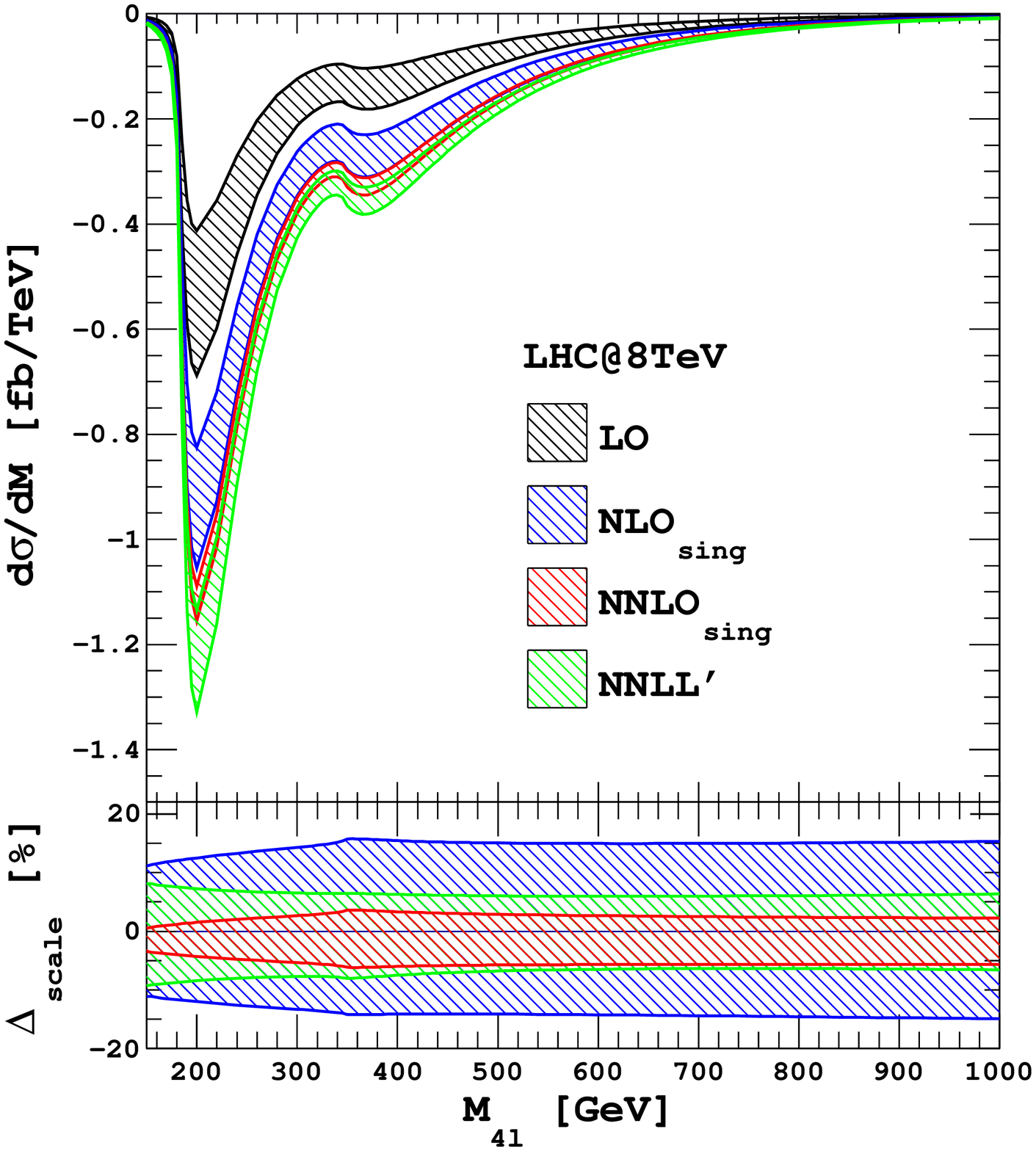}
  \includegraphics[width=0.45\linewidth]{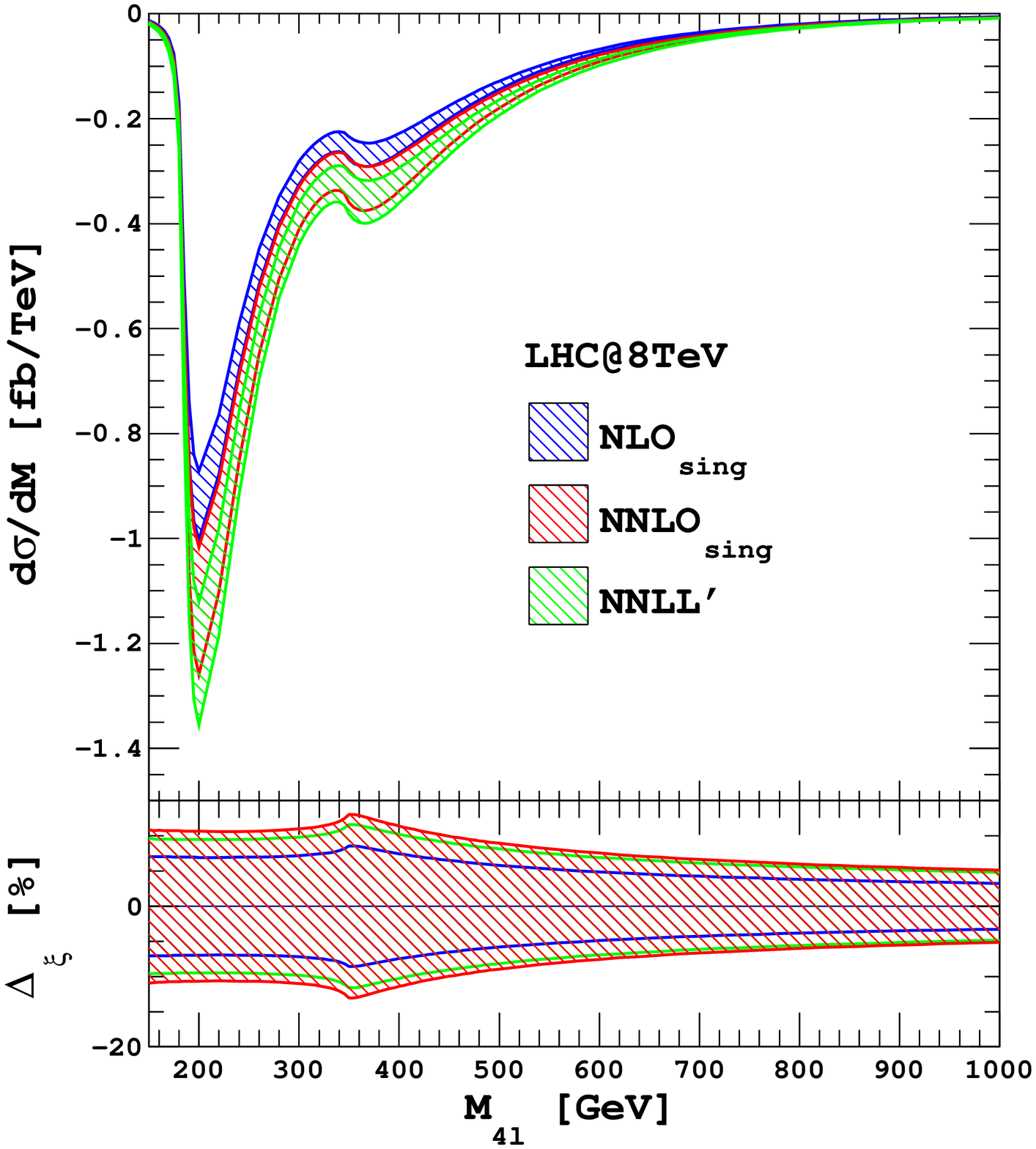}
  \\
  \caption{The cross section and uncertainties for the interference processes at the 8 TeV LHC.
  In the left plot, the LO scale uncertainties are about $\pm 20\% \sim \pm 30\%$, not shown in the plot.
  The uncertainty in the right plot refers to the uncalculated multi-loop amplitudes of the background. }
  \label{eps:int}
\end{figure}

We now provide such a theoretical prediction with higher order effects for the interference processes
between the diagrams \ref{eps:lo}(a) and  \ref{eps:lo}(b).

As shown in figure \ref{eps:int}, the contribution of the interference process is negative
\footnote{We have used the notations ${\rm NLO_{sing}}$ and ${\rm NNLO_{sing}}$  to denote the contributions from singular terms at ${\rm NLO}$ and ${\rm NNLO}$, respectively.
We keep the notations in figures and tables, but neglect the subscript in the text for simplicity. },
even overwhelming the positive pure signal,
and has a sharp peak valley around $M_{ZZ}=200$ GeV.
There is also a small valley at $M_{ZZ}=380$ GeV,
which becomes more significant at higher orders.
In contrast, the two peaks in the pure signal are of almost the same height.
We have checked that the shape of the differential cross section actually
depends on the choice of the scale and kinematical cut.
If a fixed scale, e.g., $m_h$ is used, then all the differential cross section would increase
and the cross section in the larger $M_{ZZ}$ region gets more significant improvement
compared to the case of dynamical scales.
However, since the perturbative expansion of the cross section calculated by a dynamical scale converges better,
we choose the dynamical scale in our calculation.

\begin{figure}\center
  \includegraphics[width=0.45\linewidth]{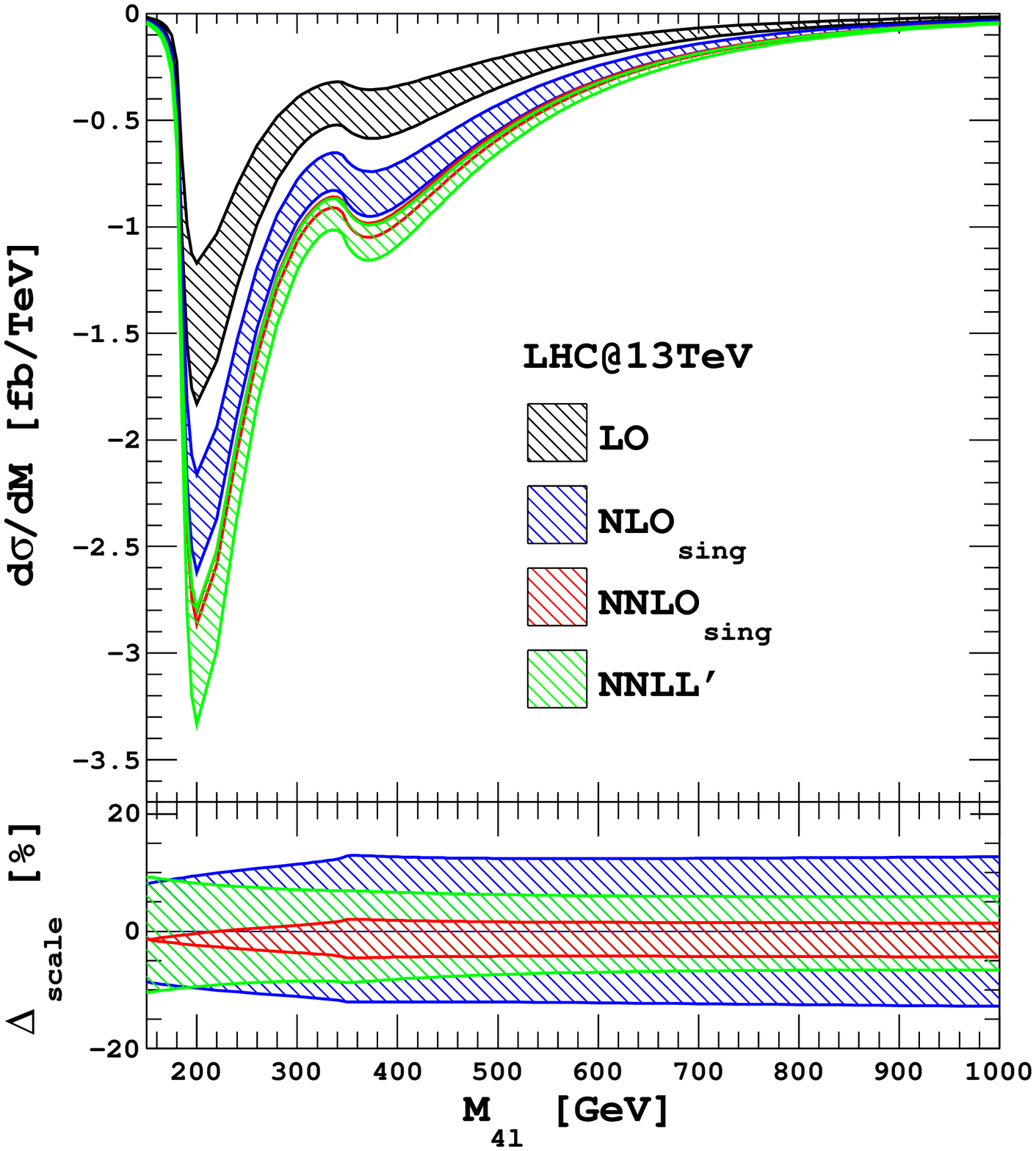}
  \includegraphics[width=0.45\linewidth]{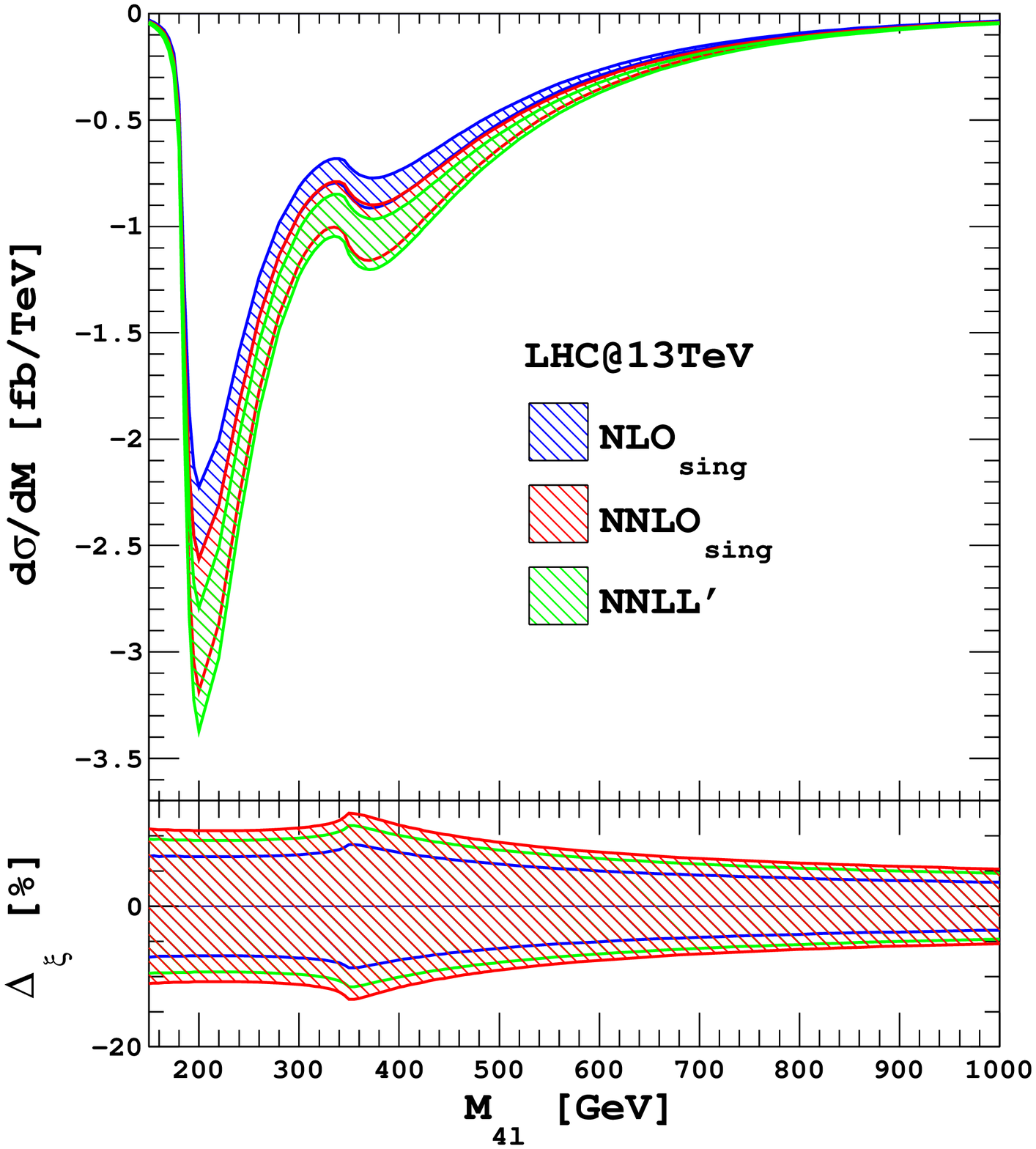}
  \\
  \caption{Same as figure \ref{eps:int}, but at the 13 TeV LHC.}
  \label{eps:int13}
\end{figure}

The results at the 13 TeV LHC are shown in figure \ref{eps:int13}.
The shapes of the differential cross sections are almost the same as at the 8 TeV LHC, as shown in figure \ref{eps:int}.

\begin{figure}\center
  \includegraphics[width=0.48\linewidth]{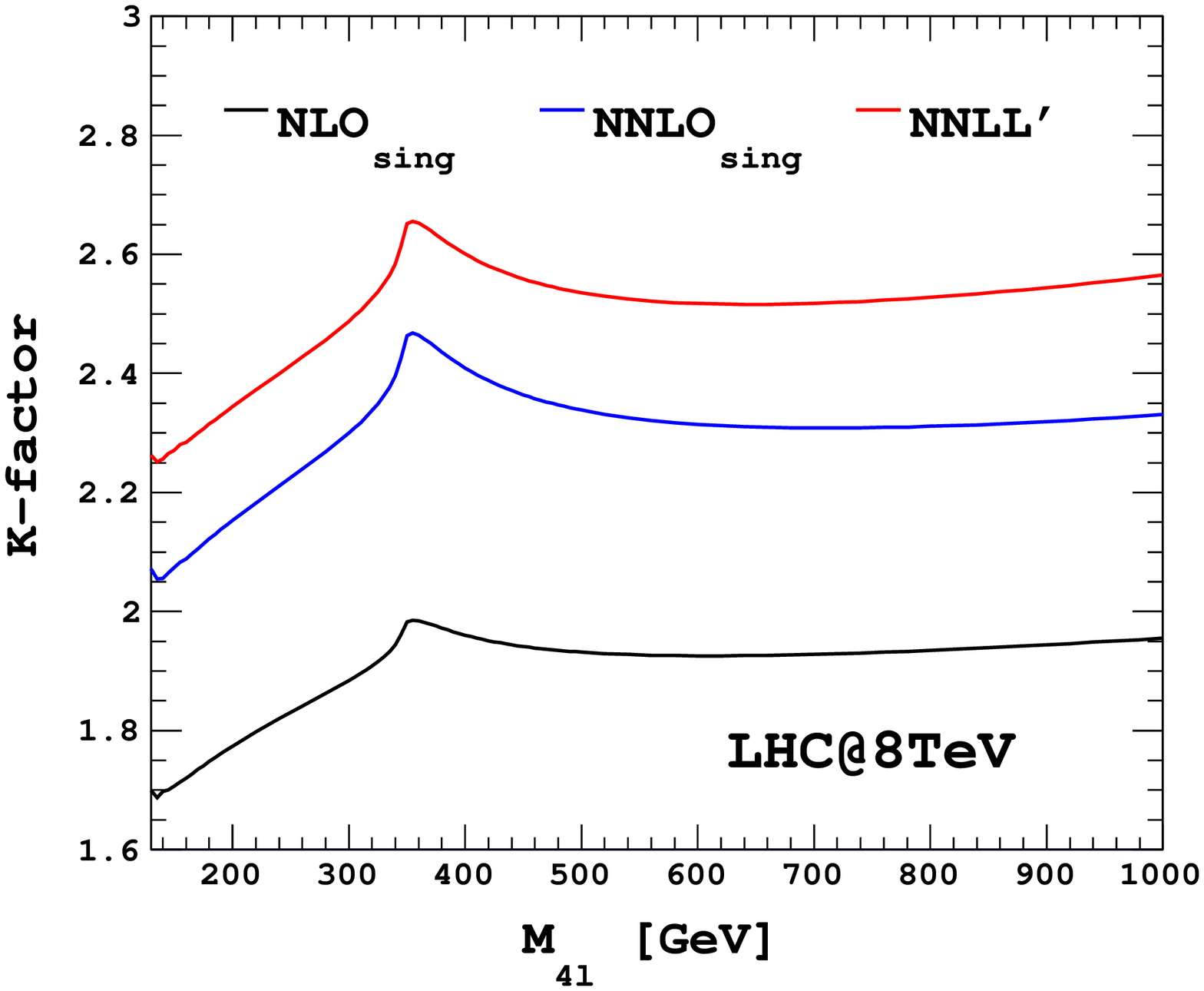}
  \includegraphics[width=0.48\linewidth]{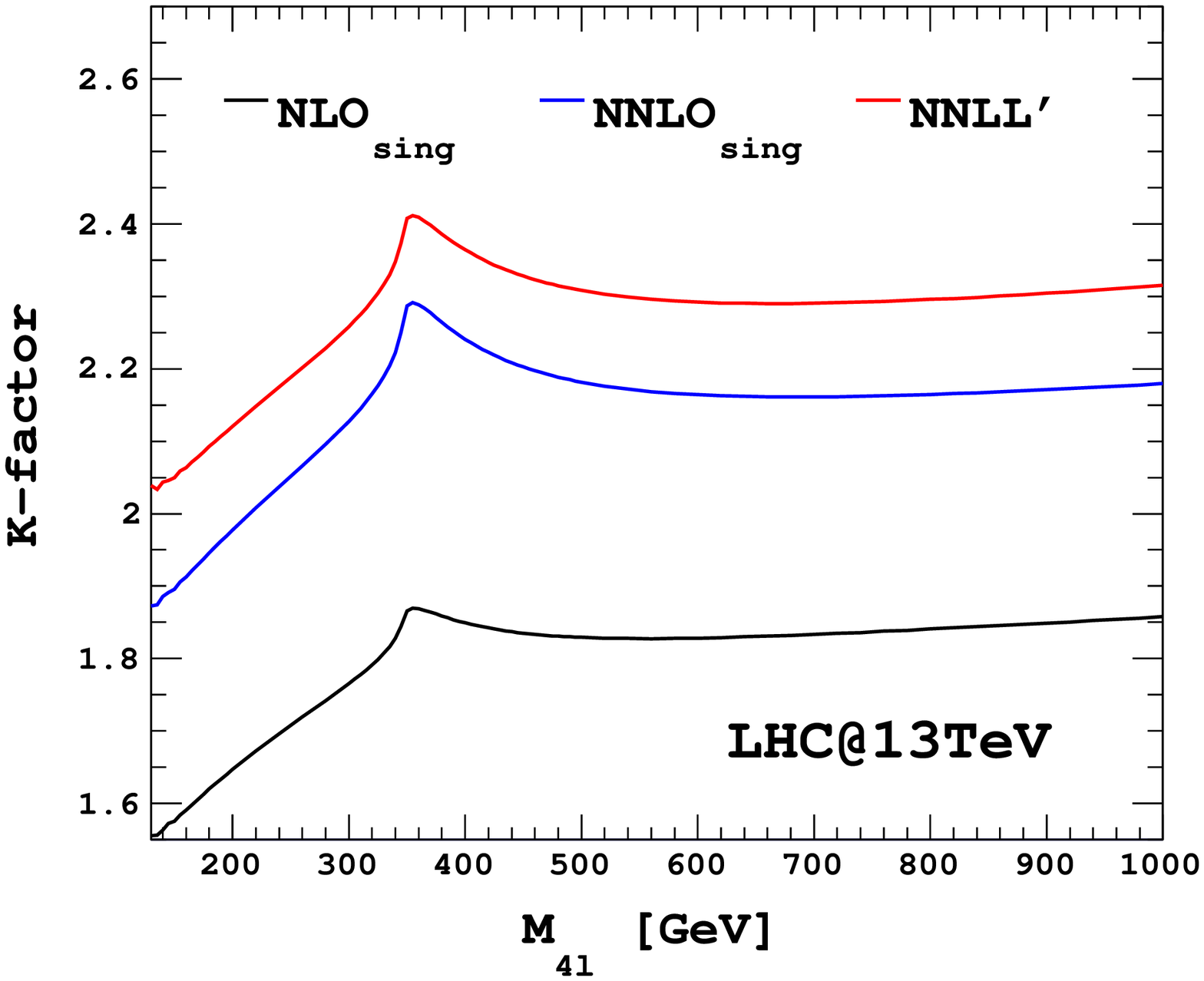}
  \\
  \caption{The  $K$-factor for the  interference process at the 8 and 13 TeV LHC.  }
  \label{eps:intK}
\end{figure}

The $K$-factors  for the  interference process at the 8 and 13 TeV LHC are shown in figure \ref{eps:intK}.
They increase from $M_{ZZ}=130$ GeV to $M_{ZZ}=350 {~\rm GeV} \approx 2m_t$,
where the interference contributions are most significant.
Then they decrease with the increasing of $M_{ZZ}$, and nearly unchanged for $M_{ZZ}>500$ GeV.
The dependences of the $K$-factor on the invariant mass at the 13 TeV LHC are similar, but the values are a little smaller.
The overall NNLO $K$-factor is in the range of $2.05-2.45$ ($1.85-2.25$) at the 8 (13) TeV LHC.
Here we point out that the ratio of $K_{\rm NNLL'}/K_{\rm NLO}$ (about $1.3$)
is similar to that in the Higgs pair production we have studied earlier \cite{Shao:2013bz}, as expected.
We also observe from figure \ref{eps:intK} that
$K_{\rm NNLL'}/K_{\rm NNLO}$ is about $1.1$
at both the 8 and 13 TeV LHC.
This means that the soft gluon resummation is important in providing more accurate theoretical predictions.

\begin{figure}\center
  \includegraphics[width=0.7\linewidth]{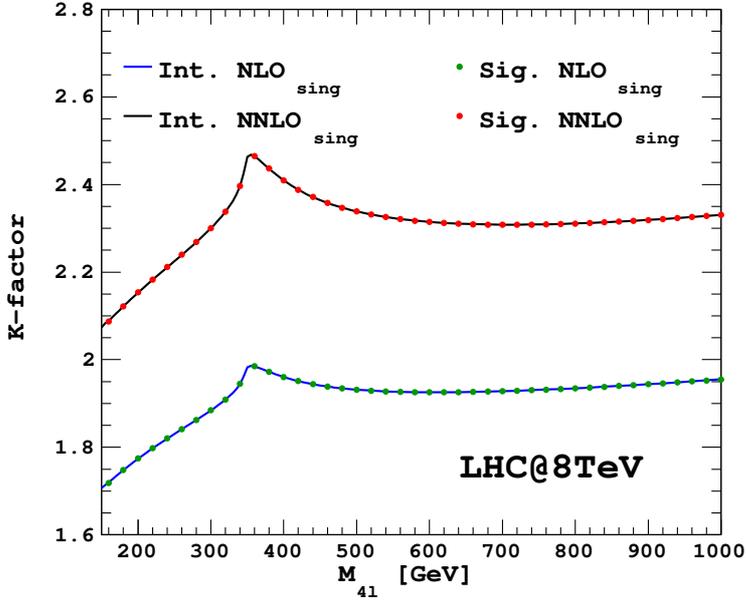}
  \\
  \caption{Comparison of the  $K$-factor in the pure signal and interference processes at the 8 TeV LHC.  }
  \label{eps:compK}
\end{figure}

Figure \ref{eps:compK} shows that the $K$-factors in the pure signal and interference processes are the same,
which is in agreement with the statements for $gg\to H(600 ~\rm GeV) \to WW$  \cite{Caola:2013yja}
and for $gg\to ZZ$ without an intermediate resonance  \cite{Melnikov:2015laa}.

\begin{table}[h!] 
\caption{The cross section for the interference process  in the region $220~{\rm GeV}<M_{ZZ}<1000$ GeV.
Besides, we also list the scale, parameter $\delta_{1,2}$ and PDF$+\as$ uncertainties.}
\centering
\begin{tabular}{lc|cc|cc|cc}
\toprule
$\sqrt{S}=8$ TeV& &\multicolumn{2}{c|}{$\Delta_{\rm scale}[\%]$}  &\multicolumn{2}{c|}{$\Delta_{\xi}[\%]$} &\multicolumn{2}{c}{$\Delta_{{\rm PDF}+\as}[\%]$}\\
\cmidrule{3-8}
LO [fb] & -0.056 & +33.7 & -23.6 & 0 & 0 & +2.5 & -2.6\\
\midrule
${\rm NLO_{sing}}$[fb] & -0.106 & +14.5 & -13.6 & +6.7 & -6.7 & +3.8 & -3.4\\
\midrule
${\rm NNLO_{sing}}$[fb] & -0.129 & +2.7 & -5.5 & +10.3 & -10.3 & +4.0 & -3.7\\
\midrule
$ {\rm NNLL'}_{{\rm w/o-\pi^2} }$[fb] & -0.110 & +10.6 & -6.6 & +7.3 & -7.3 & +3.8  & -3.5 \\
\midrule
${\rm NNLL'}$[fb] & -0.140 & +7.4 & -6.4 & +9.3 & -9.3 & +4.3 & -3.9\\
\bottomrule
\end{tabular}
\label{tab:unce8}
\end{table}

\begin{table}[h!] 
\caption{Same as table \ref{tab:unce8}, but at 13 TeV LHC.}
\centering
\begin{tabular}{lc|cc|cc|cc}
\toprule
$\sqrt{S}=13$ TeV& &\multicolumn{2}{c|}{$\Delta_{\rm scale}[\%]$}  &\multicolumn{2}{c|}{$\Delta_{\xi}[\%]$} &\multicolumn{2}{c}{$\Delta_{{\rm PDF}+\as}[\%]$}\\
\cmidrule{3-8}
LO [fb] & -0.189 & +29.4 & -21.4 & 0 & 0 & +1.9 & -2.2\\
\midrule
${\rm NLO_{sing}}$[fb] & -0.339 & +11.8 & -11.5 & +6.7 & -6.7 & +3.2 & -2.6\\
\midrule
${\rm NNLO_{sing}}$[fb] & -0.407 & +1.2 & -3.9 & +10.2 & -10.2 & +3.4 & -3.0\\
\midrule
$ {\rm NNLL'}_{\rm w/o-\pi^2}$[fb] & -0.340 & +10.3 & -7.6 & +6.7 & -7.1 &  +3.2 & -2.8 \\
\midrule
${\rm NNLL'}$[fb] & -0.432 & +7.8 & -6.8 & +8.9 & -8.9 & +3.7 & -3.2\\
\bottomrule
\end{tabular}
\label{tab:unce13}
\end{table}

Next, we discuss the theoretical uncertainties in the results about the interference process,
which have been shown in the bottom plots in figures \ref{eps:int} and  \ref{eps:int13}.
The scale uncertainties at LO, NLO and NNLO are obtained by varying $\mu_r=\mu_f$
in the range of [$M_{ZZ}$/2, $2M_{ZZ}$].
For the results at ${\rm NNLL}'$, we first vary $\mu_f$, $\mu_s$ and $\mu_h$ independently by a factor of 2,
and then combine the errors in quadrature.
The LO scale uncertainties are $\pm 20\%\sim \pm 30\%$,
which are so large that we do not show them out in the figures.
At NLO, NNLO and ${\rm NNLL}'$, they are about $\pm 15\%$, $\pm 5\%$ and  $\pm 6\%$, respectively.
Therefore the scale uncertainties are significantly reduced after including higher order QCD corrections.
The uncalculated multi-loop amplitudes of the background are evaluated
by changing the parameter $\xi$ from 0 to 2, as described at the beginning of this section.
The numerical results are also shown in figures \ref{eps:int} and \ref{eps:int13}.
The associated theoretical uncertainties are about $5\%-10\%$, depending on $M_{ZZ}$,
and slightly increase from NLO to NNLO and  ${\rm NNLL}'$.

In tables \ref{tab:unce8} and \ref{tab:unce13}, we list the PDF$+\alpha_s$ uncertainties
for the interference process  in the region $220~{\rm GeV}<M_{ZZ}<1000$ GeV.
They are at most about $4\%$, much less than the other uncertainties.
If all the theoretical uncertainties are added in quadrature,
the uncertainties at ${\rm NNLL}'$ are about $\pm 12\%$. The results at the 13 TeV LHC are similar.
For comparison, we also list the resummation results without $\pi^2$ enhanced terms
by setting the hard scale $\mu_h^2=M_{ZZ}^2$ in eq.(\ref{eqs:hardWilson}), denoted by ${\rm NNLL'}_{\rm w/o-\pi^2}$.
We find that this kind of resummed cross section is close to the NLO singular terms, much less than the NNLO singular and ${\rm NNLL}'$ results, which means that the $\pi^2$-enhanced terms make the main contributions  to the ${\rm NNLL}'$ results.
This feature is in agreement with that in single Higgs production \cite{Ahrens:2008nc} and the double Higgs  production \cite{Shao:2013bz}  at the LHC.
The scale uncertainty at ${\rm NNLL'}_{\rm w/o-\pi^2}$ is reduced compared to the NLO singular terms,
but a little larger than total resummed result.
The other theoretical uncertainties at ${\rm NNLL'}_{\rm w/o-\pi^2}$ are nearly the same as the NLO singular terms.

\begin{figure}\center
  \includegraphics[width=0.55\linewidth]{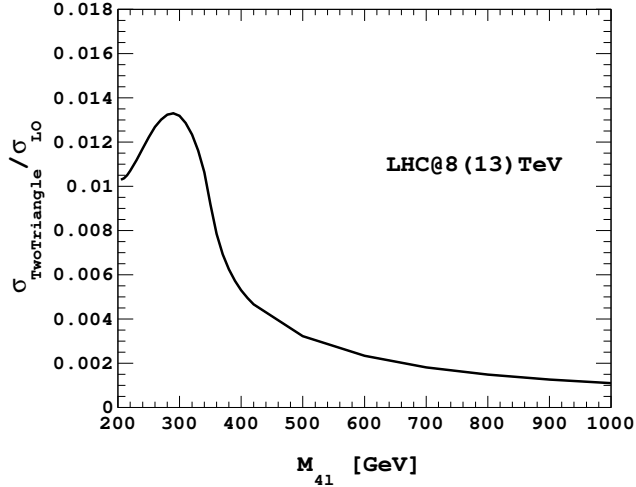}\\
  \caption{ The ratio of the contribution from the interference between the diagrams \ref{eps:lo}(a) and \ref{eps:nlo}(c)
   to the LO cross section.  }
  \label{eps:twotriangle}
\end{figure}

Now we evaluate a special contribution in $gg(\to h^*)\to ZZ$ but not in $gg(\to H)\to WW$, i.e.,
the interference between the diagrams \ref{eps:lo}(a) and \ref{eps:nlo}(c),
which would appear at the NLO corrections.
Furry's theorem states that only the axial vector part of the $Z$ boson coupling can contribute.
Since the coefficient of the axial current is proportional to the weak charge $T_{w3}$ of the $SU(2)_L$ gauge group,
the contribution is proportional to the difference of $T_{w3}$ of the quarks in a $SU(2)_L$ doublet.
Thus only the third generation quarks, massive t- and b-quarks, generate non-vanishing result,
which is both infrared and ultraviolet safe.
As shown in figure \ref{eps:twotriangle}, they are so small that we can neglect them when considering
higher order corrections.

\section{Conclusion}
\label{sec:conclusion}

We have studied the high order QCD effects, in particular the soft gluon resummation, in the
signal-background interference process of $gg(\to h^*) \to ZZ$.  This process can be used
to constrain the total width of the Higgs boson and provide a special way to measure
the couplings between the Higgs boson and other particles in SM.
The previous theoretical prediction for the interference process is only at the LO because of
technical difficulty in calculating massive two-loop integrals.
In this work we show approximate NLO and NNLO cross sections obtained from the resummation formalism in SCET,
and also present the result with the soft gluon effect resummed to all orders in $\alpha_s$.
Comparing the approximate  results with the exact ones for the pure signal process at NLO and NNLO,
we observe that approximate results almost reproduce the exact ones in the off-shell region,
and find that the high order QCD effects for the interference process are very sizable.
For the interference process, the approximate NNLO $K$-factor is in the range of $2.05-2.45$ ($1.85-2.25$),
depending on $M_{ZZ}$, at the 8 (13) TeV LHC.
Besides, the soft gluon resummation can increase the approximate NNLO result by about $10\%$ at both the 8 TeV and 13 TeV LHC.
At the same time,  the corrections from the soft gluon resummation are
similar to that in the Higgs pair production.
We also find that the approximate $K$-factors in the interference and the pure signal processes are the same,
which is in agreement with the statements in previous literatures \cite{Caola:2013yja,Melnikov:2015laa}.

Moreover, we discuss  the theoretical uncertainties in the results.
The scale uncertainties are significantly reduced after including higher order QCD corrections.
The uncertainties from uncalculated multi-loop amplitudes of the background slightly increase
from NLO to NNLO and  ${\rm NNLL}'$.
The PDF$+\alpha_s$ uncertainties are rather small compared with the others.
Combined all together, the theoretical predictions at ${\rm NNLL}'$ suffer from roughly $\mathcal{O}(10\%)$ uncertainties.

Our study can be easily extended to other processes, such as the pure signal process $gg\to h^* \to VV$
and pure background process $gg\to VV ~(V=W,Z)$ via quark box-loops.

\section{Acknowledgement}

We would like to thank Thomas Becher for carefully reading our manuscript and providing helpful comments.
We are grateful to Fabrizio Caola for explaining the details of their work.
C.S. Li and H.T. Li was supported in part by the National Natural Science
Foundation of China under Grants No.11375013 and No.11135003.
This work of D.Y. Shao
is supported by the Swiss National Science Foundation (SNF) under the  grant
number 200020\_153294.
The research of J. Wang has been supported by
the Cluster of Excellence {\it Precision Physics, Fundamental Interactions and Structure of Matter} (PRISMA-EXC 1098).
The Feynman diagrams in the paper are drawn by {\tt Jaxodraw} \cite{Binosi:2003yf}.

\bibliography{hzz}
\bibliographystyle{JHEP}

\end{document}